\documentclass[conference]{IEEEtran}

\usepackage{amsmath,amsfonts,amssymb,mathtools}
\usepackage{graphicx}
\usepackage{booktabs}
\usepackage{multirow}
\usepackage{tabularx}
\usepackage{makecell}
\usepackage{enumitem}
\usepackage{url}
\usepackage[hidelinks]{hyperref}
\usepackage{algorithm}
\usepackage{algpseudocode}
\usepackage{tikz}
\usepackage{caption}
\usetikzlibrary{arrows.meta,positioning,shapes.geometric,fit,calc,backgrounds}

\usepackage[most]{tcolorbox}

\tcbset{
    insightbox/.style={
        colback=gray!10,
        colframe=black,
        fonttitle=\bfseries,
        title=Key Insight,
        boxrule=0.8pt,
        arc=2pt,
        left=6pt,
        right=6pt,
        top=6pt,
        bottom=6pt,
        breakable        
    }
}

\begin{document}

\title{Radar Detection in the CBRS Band: Techniques, Challenges, and Future Directions}

\author{%
  \IEEEauthorblockN{Madan Baduwal, Priyanka Paudel}
  \IEEEauthorblockA{%
    \textit{Department of Computer Science and Engineering,}\\
    \textit{Mississippi State University,}
    Mississippi State, MS, USA\\
    \{mb4239, pp918\}@msstate.edu}
}

\maketitle

\begin{abstract}
The $3.5$~GHz Citizens Broadband Radio Service (CBRS) is a shared wireless band that allows both government systems and commercial networks (such as private LTE/5G) to use the same spectrum. To prevent interference with critical government systems, especially naval radars, CBRS uses a monitoring system called the Environmental Sensing Capability (ESC). ESC acts like a network of sensors that continuously listens for radar signals and alerts the system when they are detected, so commercial users can temporarily stop or adjust their transmissions. This paper reviews how radar signals are detected within the CBRS band. We first explain the regulatory framework and describe the types of radar signals that need to be identified. We then examine traditional detection methods, such as energy-based and pattern-matching techniques, and compare them with newer approaches based on machine learning and deep learning, which can automatically learn to recognize radar signals from data. We also review publicly available datasets and testing platforms used to evaluate these detection methods, along with key performance requirements such as high detection accuracy (e.g., $99\%$ detection probability (radar overlap recall)) and low delay (e.g., within $60$~seconds). Finally, we highlight current challenges, including false alarms, interference from modern wireless systems, and the need for real-time operation. Overall, this survey shows that while traditional methods are simple and reliable in controlled settings, modern learning-based approaches offer better performance in complex environments. The future of CBRS radar detection will likely combine both approaches to achieve accurate, fast, and robust performance in real-world deployments.
\end{abstract}

\section{Introduction}

The Citizens Broadband Radio Service (CBRS) band (around $3.45$--$3.7$~GHz) is a special part of the wireless spectrum in the United States that is shared between government systems and commercial users like private LTE and $5$G networks. Instead of giving this spectrum to only one group, the Federal Communications Commission (FCC) designed a system that allows multiple users to safely share it.

\begin{figure*}[!t]
  \centering
  \includegraphics[width=0.75\textwidth]{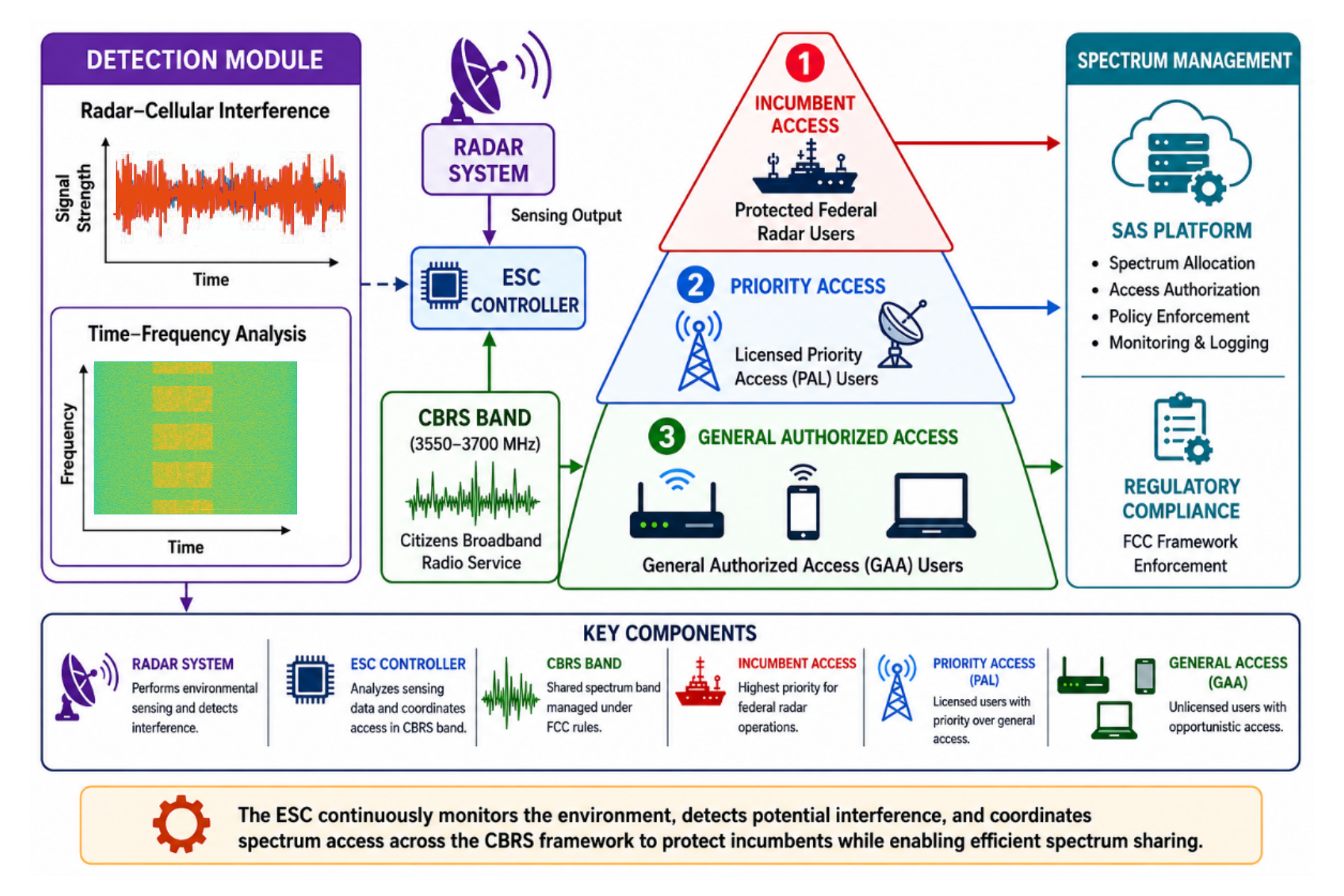}
  \caption{CBRS Spectrum Sharing Architecture and ESC Operation.
The three-tier access hierarchy — Incumbent (federal radar), Priority Access (PAL), and General Authorized Access (GAA) is managed by the Spectrum Access System (SAS). ESC sensors continuously monitor the $3550-3700$ MHz band, detect naval radar signals, and coordinate with the SAS to enforce spectrum evacuation and regulatory compliance.}
  \label{fig:cbrs}
\end{figure*}

To manage this sharing, CBRS uses a three-level access system. At the top level are government users, such as U.S.\ Navy radar systems, which must always be protected. Below them are licensed commercial users, and finally general users who can access the spectrum when it is available. This system is controlled by a central manager called the Spectrum Access System (SAS). A key challenge in this shared system is making sure that commercial users
do not interfere with important government radar systems. To solve this, CBRS uses something called the Environmental Sensing Capability (ESC). ESC is a network of sensors, usually placed near coastlines, that continuously monitor the spectrum and listen for radar signals. When a radar signal is detected, the ESC quickly informs the SAS, which then tells commercial systems to stop or adjust their transmissions to avoid interference~\cite{FCC,WInnForum1}. The FCC has strict requirements for this detection system. For example, the ESC must detect radar signals with at least a $99\%$ probability (radar overlap recall) within $60$ seconds (to enable timely detection and vacating), even when the signal is extremely weak (as low as $-89$,dBm/MHz)~\cite{FCC2015}. Meeting these requirements is
difficult because radar signals can vary in shape, strength, and timing, and they may overlap with signals from modern wireless systems like LTE and Wi-Fi. The overall CBRS architecture and the interaction between ESC sensors and the SAS are illustrated in Fig.~\ref{fig:cbrs}.

Because of these challenges, researchers have developed different methods to detect radar signals. Traditional methods rely on signal processing techniques, such as measuring signal energy or matching known radar patterns. These methods are simple and reliable in controlled environments but may struggle in real-world conditions with noise and interference. More recently, machine learning and deep learning approaches have been introduced. These methods can automatically learn patterns from data and often perform better in complex environments.

This paper provides an overview of radar detection techniques in the CBRS band. It explains how the system works, describes the types of radar signals involved, and compares traditional and modern detection methods. It also discusses datasets, evaluation metrics, and real-world challenges. Finally, the paper highlights open research problems and future directions in this area.

\noindent\textbf{Contributions of this Survey:}
\begin{itemize}
\item A unified taxonomy and system-level pipeline integrating classical and learning-based radar detection methods.
\item A comparative analysis of detection techniques under CBRS-specific constraints such as low-SNR operation and strict latency requirements.
\item A comprehensive review of datasets, simulation platforms, and evaluation benchmarks used in ESC validation.
\item Identification of key deployment challenges and emerging hybrid detection strategies.
\item Discussion of open research problems and future directions for scalable and adaptive CBRS sensing systems.
\end{itemize}

The rest of the paper is organized as follows. 
Section~\ref{sec:related-work} reviews previous research on radar detection methods. 
Section~\ref{sec:methodology_literature_search} explains how the literature was selected. 
Section~\ref{sec:cbrs_and_ESC_background} describes the CBRS system and ESC in more detail. 
Section~\ref{sec:radar_signal_charastics_in_the_cbrs_band} discusses radar signal characteristics. 
Sections~\ref{sec:Classical_Signal_Processing_Detection_Methods}, 
\ref{sec:Machine-Learning_and_Deep-Learning_Approaches}, and 
\ref{sec:Hybrid_Detection_Approaches} cover classical, machine-learning, and hybrid detection methods, respectively.
Section~\ref{sec:Datasets_and_Simulation} reviews available datasets, and 
Section~\ref{sec:Performance_Metrics_and_Benchmarks} discusses performance metrics. 
Section~\ref{sec:Implementation_and_Deployment_Challenges} explains implementation challenges, followed by spectrum sharing strategies in 
Section~\ref{sec:Spectrum_Sharing_Implications_and_Mitigation_Strategies}. 
Section~\ref{sec:Open_Research_Problems_and_Future_Directions} highlights future research directions, and 
Section~\ref{sec:Conclusions} concludes the paper.

\section{Related Work}
\label{sec:related-work}

\noindent\textbf{Radar Detection in the CBRS Band:} Over time, the way we detect radar signals in the CBRS band has improved a lot. Early approaches mainly used traditional signal processing techniques. One common method was the matched-filter (MF) detector, which works by comparing incoming signals with known radar patterns~\cite{lees2019deep}. These methods work well when the radar signal is already known and conditions are simple. However, they are not very flexible and may fail when signals change or when there is strong interference. To improve this, researchers started using machine learning methods. These methods use features (important characteristics of signals) and classifiers such as Support Vector Machines (SVMs)~\cite{8751641}. While these approaches are more robust than older methods, they still depend heavily on manually designed features and labeled training data, which can limit their performance in real-world situations.

More recently, deep learning methods have become very popular for radar detection. These methods can automatically learn patterns directly from data, such as spectrogram images or raw IQ signals. Models like Convolutional Neural Networks (CNNs) and other deep architectures~\cite{krizhevsky2012imagenet,he2016deep,Caromi2021,lees2019deep} have shown very strong performance across different scenarios. For example, ESC+~\cite{soltani2022finding} uses a YOLO-based model to detect radar signals from spectrograms and achieves over $99\%$ accuracy at $17$\,dB SINR. Another system called DeepRadar~\cite{DeepRadar} also reports similar performance when the signal quality is above $20$\,dB. Other systems such as Spec-SCAN~\cite{10976013} and SenseORAN~\cite{senseORAN} show that deep learning models can still perform well even when signals are weak or noisy.

Some researchers also use raw IQ data instead of spectrograms~\cite{Caromi2021}. This helps keep more detailed information about the signal, such as phase and timing. However, these methods can be more complex and may be sensitive to interference.

Recently, researchers have started combining multiple types of inputs to improve detection accuracy. For example,~\cite{Shafi2025} combines both IQ data and spectrograms in one model and achieves nearly $99\%$ accuracy even in high-interference environments. In addition, techniques like model compression and quantization help reduce computation cost so these models can run faster in real systems.

Existing works can be broadly categorized into four groups: (i) classical signal-processing-based methods~\cite{lees2019deep}, (ii) feature-based machine learning approaches~\cite{8751641}, (iii) deep learning-based methods~\cite{Lees2019,DeepRadar}, and (iv) hybrid approaches that combine classical and learning-based techniques~\cite{Shams2026}. Classical methods offer theoretical guarantees but lack adaptability in heterogeneous environments, while feature-based machine learning approaches improve robustness yet remain dependent on manually engineered signal representations. Deep learning methods provide superior detection performance in complex, interference-heavy environments, often exceeding 99\% accuracy, at the cost of higher computational requirements and reliance on large labeled datasets. Hybrid approaches address these limitations by integrating the low-latency reliability of classical detectors, such as CFAR, with the pattern recognition capability of deep learning models, achieving strong performance while reducing false alarms under low-SNR conditions. Despite these advances, most current radar detection systems are trained in controlled environments using labeled data and may not generalize well to real-time deployments where signal conditions change continuously. Therefore, making these models more adaptive, efficient, and suitable for real-world ESC deployment remains an important and open research challenge.

\section{Methodology of Literature Search}
\label{sec:methodology_literature_search}


\begin{figure}[!t]
    \centering
    \includegraphics[width=\columnwidth]{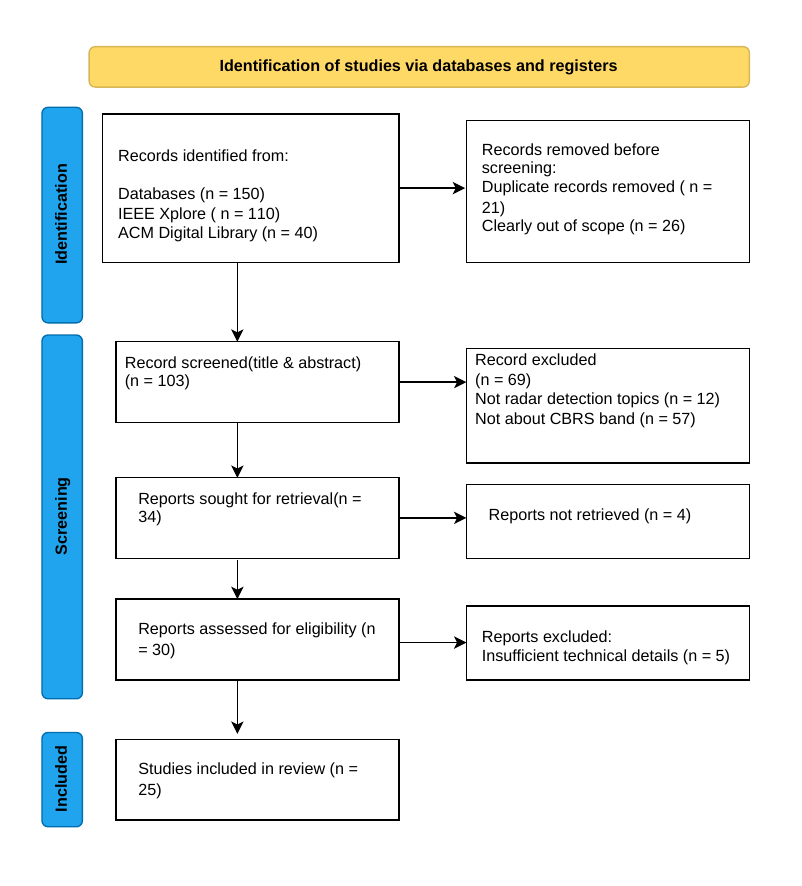}
    \caption{PRISMA flow diagram illustrating the study selection process for radar detection in the CBRS band. A total of 150 records were initially identified from IEEE Xplore and ACM Digital Library, with 27 studies ultimately included after screening and eligibility assessment.}
    \label{fig:prisma}
\end{figure}

This survey follows a structured literature review process guided by the PRISMA framework, as illustrated in Fig.~\ref{fig:prisma}. The objective was to systematically identify, screen, and select relevant studies on radar detection in the CBRS band published between 2018 and 2026. The identification phase began with a comprehensive search of major digital libraries, primarily IEEE Xplore and the ACM Digital Library. A total of 150 records were retrieved using targeted keywords such as ``CBRS radar detection,'' ``3.5~GHz spectrum sensing,'' ``ESC,'' ``SAS,'' and ``machine learning for radar detection.'' In addition to academic publications, official regulatory documents from the Federal Communications Commission (FCC), particularly 47 CFR Part 96~\cite{FCC}, and technical specifications from the Wireless Innovation Forum (WInnForum)~\cite{WInnForum1,WInnForum2} were included to capture system-level requirements and real-world deployment considerations.

During the screening phase, duplicate entries (n = 21) and clearly irrelevant studies (n = 26) were removed, resulting in 103 records for title and abstract screening. Of these, 69 records were excluded due to lack of relevance, including studies not focused on radar detection (n = 12) or not related to the CBRS band (n = 57). The remaining 34 reports were considered for full-text retrieval, of which 4 could not be accessed. In the eligibility assessment stage, 30 full-text articles were carefully evaluated based on technical relevance, methodological clarity, and contribution to the field. Studies lacking sufficient technical depth (n = 5) were excluded. This process resulted in a final set of 25 studies included in this survey. To ensure comprehensive coverage, the selected studies span a diverse range of approaches, including traditional signal processing techniques such as matched filtering~\cite{Caromi2018}, as well as modern data-driven methods based on machine learning and deep learning~\cite{Lees2019}. Additionally, experimental platforms and publicly available datasets, including NIST measurement campaigns~\cite{Hale2017,NISTDataset2020} and recent O-RAN-based implementations~\cite{ReusMuns2023}, were reviewed to incorporate practical evaluation perspectives.

Overall, this systematic methodology ensures a balanced and thorough review by integrating regulatory guidelines, theoretical advancements, and experimental validations. The resulting dataset of selected studies provides a solid foundation for analyzing detection techniques, system architectures, performance metrics, and open research challenges in the CBRS radar detection domain.
\section{CBRS and ESC Background}
\label{sec:cbrs_and_ESC_background}

The Citizens Broadband Radio Service (CBRS) operates around the $3.5$~GHz frequency band, originally defined as $3550$--$3700$~MHz and later expanded to include $3450$--$3550$~MHz~\cite{FCC2015}. This band is special because it is shared between government systems and commercial users instead of being assigned to only one group. To make this sharing work safely, the Federal Communications Commission (FCC) created a three-level access system~\cite{FCC}:

\begin{itemize}
    \item \textbf{Tier 1 (Incumbent Access):} Government users such as
    naval radar systems. These users have the highest priority and must
    always be protected.
    
    \item \textbf{Tier 2 (Priority Access License):} Licensed commercial
    users who have reserved access to parts of the spectrum.
    
    \item \textbf{Tier 3 (General Authorized Access):} General users who
    can use the spectrum when it is available.
\end{itemize}

A central system called the Spectrum Access System (SAS) manages this
sharing by deciding who can use which frequency and when~\cite{WInnForum1}. One of the biggest challenges in CBRS is protecting government radar systems from interference caused by commercial networks. To solve this, CBRS uses the Environmental Sensing Capability (ESC). The ESC is a network
of sensors that continuously monitor the radio signals in the CBRS band. These sensors are usually placed near coastlines because many naval radars operate in those areas. The main job of the ESC is to detect radar signals in real time. According to FCC rules, the ESC must accurately detect signals from federal systems in the $3550$--$3700$~MHz band~\cite{FCC}. When a radar signal is detected,
the ESC quickly sends this information (such as frequency and timing) to the SAS. The SAS then tells nearby commercial systems to stop or reduce their transmissions so they do not interfere with the radar. The Wireless Innovation Forum (WInnForum) provides technical details about how ESC systems should work~\cite{WInnForum1,WInnForum2}. For example:
\begin{itemize}
    \item ESC sensors typically monitor frequencies between $3550$ and
    $3650$~MHz.
    
    \item Sensors are often designed to face the ocean to better detect
    naval radar signals.
    
    \item The system must handle interference carefully, with limits such
    as an aggregate interference level of about $-109$\,dBm/MHz at the
    sensor input~\cite{WInnForum2}.
\end{itemize}

The FCC also sets strict performance requirements for ESC systems. For
example:

\begin{itemize}
    \item The system must detect radar signals with at least $99\%$
    probability.
    
    \item Detection and evacuation must happen within $60$ seconds.
    
    \item The system must detect very weak signals, as low as
    $-89$\,dBm/MHz~\cite{FCC2015}.
\end{itemize}

These requirements are difficult to meet because radar signals can vary a lot in their shape, strength, and timing. Most radar signals in the CBRS band come from ship-based systems such as
air traffic control and surface search radars (for example, AN/SPN-43 and AN/SPS-67). These radars usually transmit signals in the form of short pulses or frequency sweeps (chirps)~\cite{Hale2017,Caromi2019}. Studies have shown that these radar signals can vary widely. For example, one measurement campaign collected about $12{,}000$ scans and more than
$223{,}000$ individual radar pulses at around $3570$~MHz~\cite{Caromi2019}. These pulses had very different strengths (spanning more than $40$~dB) and different timing patterns.

Even though radar signals vary, they still have recognizable patterns, such as repeating pulses or specific frequency ranges. Detection systems use these patterns to identify radar signals.

\section{Radar Signal Characteristics in the CBRS Band}
\label{sec:radar_signal_charastics_in_the_cbrs_band}


Radar signals in the CBRS band may seem complex at first, but they can be understood using a few simple concepts. Most radar systems do not transmit continuously; instead, they send signals in short bursts called \textit{pulses}. In some cases, radars use \textit{chirp} signals, where the frequency changes over time. These basic signal types form the foundation of how radar operates in this band. For example, the U.S.\ Navy AN/SPN-43 radar transmits very short pulses, typically about $1$--$5$ microseconds long. These pulses repeat many times per second, usually a few hundred times per second, which is referred to as the pulse repetition frequency (PRF). The signal also occupies a bandwidth of about $1.6$~MHz~\cite{Hale2017}. These parameters provide a general idea of how radar signals behave, but they are not fixed and can vary significantly in real-world conditions.

Measurements collected from real environments show that radar signals are highly variable. In one study, more than $223{,}000$ radar pulses were recorded around $3570$~MHz~\cite{Caromi2019}. The results showed that signal strength can vary by more than $40$~dB, pulse timing is not always consistent, and the rotation of radar antennas causes signals to appear periodically. Because of this variability, radar signals are not always predictable, even though they follow general patterns. Different radar systems, such as SPN-50 and SPS-67, may use similar pulse-based or chirp-based transmissions, but their exact characteristics differ. This diversity makes it difficult to design a single detection method that works well for all radar types. Another important challenge is interference. Signals from nearby frequency bands (below $3550$~MHz) can leak into the CBRS band, making it harder to distinguish true radar signals from other sources~\cite{Caromi2018}. In practice, ESC sensors focus on monitoring frequencies between $3550$ and $3650$~MHz, where most CBRS radar activity occurs.

\begin{figure*}[!t]
  \centering
  \includegraphics[width=0.95\textwidth]{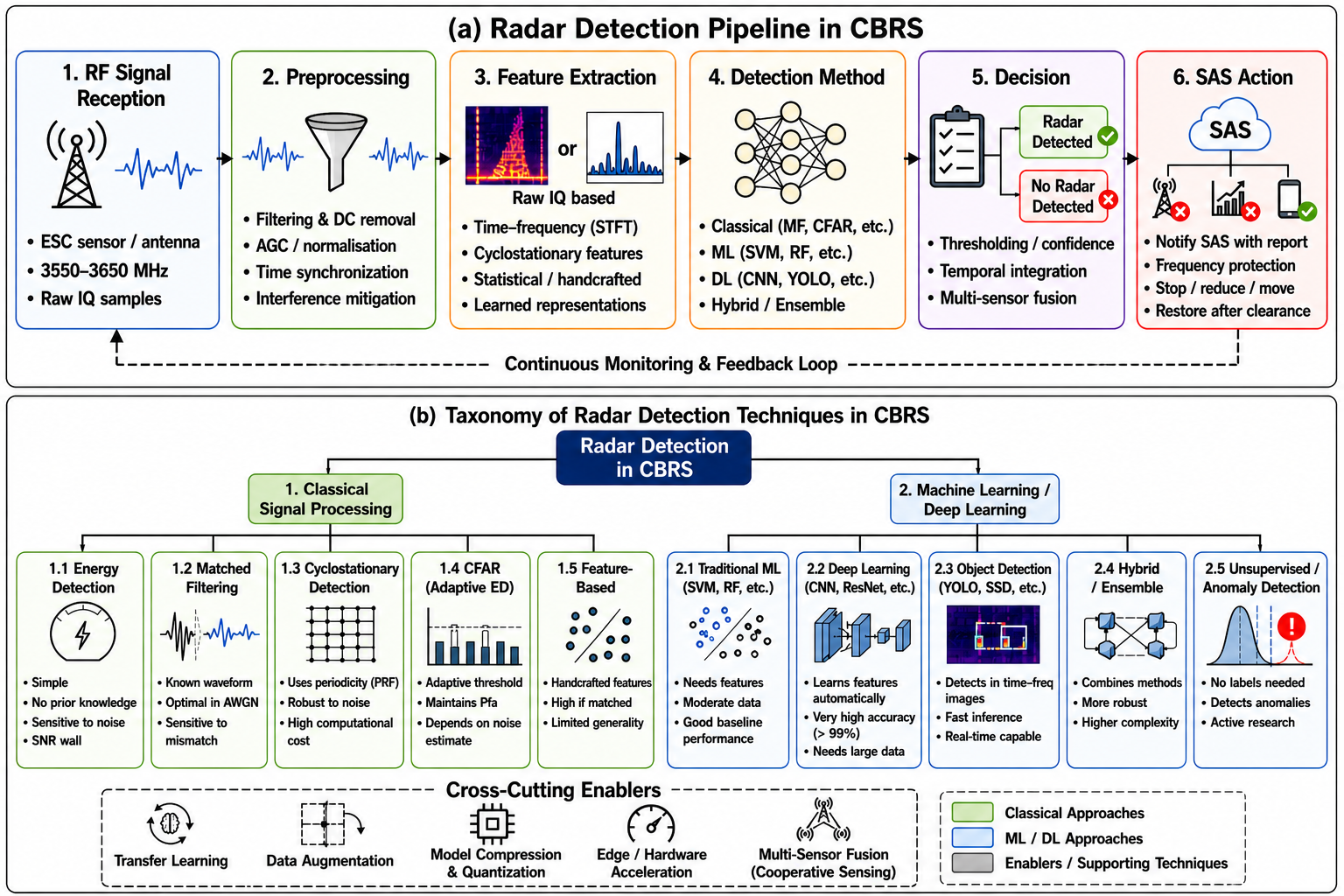}
  \caption{Overview of radar detection in the CBRS band. 
  (a) End-to-end detection pipeline from RF signal reception to SAS action, 
  illustrating key processing stages. 
  (b) Taxonomy of radar detection techniques, categorized into classical 
  signal processing and machine-learning-based approaches, along with 
  supporting enablers.}
  \label{fig:pipeline_taxonomy}
\end{figure*}

Even with these challenges, radar signals still exhibit useful patterns. Many signals repeat at regular intervals due to PRF, appear stronger when the antenna is directed toward the sensor, and may show distinct frequency patterns, especially in the case of chirps. These repeating and structured behaviors are known as \textit{cyclostationary properties}, and they help detection systems identify radar signals even when they are weak or affected by noise. Overall, detecting radar signals in the CBRS band is challenging because the signals are short, non-continuous, and highly variable in both timing and strength. They may also overlap with LTE, $5$G, or Wi-Fi signals, and interference from nearby bands can further complicate detection. Because of these factors, detection systems must be designed to handle different signal types, operate reliably in noisy environments, and detect weak signals accurately.
These variations in radar signal properties are quantitatively summarized in Table~\ref{tab:radar_params}.
\begin{table}[!t]
  \centering
  \caption{Representative Parameters of Incumbent Radar Signals in the
           3.5~GHz CBRS Band~\cite{Hale2017,Caromi2019}.}
  \label{tab:radar_params}
  \renewcommand{\arraystretch}{1.15}
  \setlength{\tabcolsep}{4pt}
  \begin{tabular}{lll}
    \toprule
    \textbf{Parameter} & \textbf{Typical Range} & \textbf{Notes} \\
    \midrule
    Center frequency  & 3550--3650~MHz  & In-band CBRS portion \\
    Pulse width       & 1--5~\textmu s  & Varies by radar type \\
    PRF               & 100--1000~Hz    & Pulse repetition frequency \\
    Bandwidth         & $\sim$1.6~MHz   & SPN-43 example \\
    Peak power range  & $>$40~dB spread & Site-dependent \\
    Antenna scan      & 2--12~s/rev     & Rotation period \\
    \bottomrule
  \end{tabular}
\end{table}
Figure~\ref{fig:pipeline_taxonomy} provides a unified overview of radar detection in the CBRS band from both a system and methodological perspective. The pipeline view in Fig.~\ref{fig:pipeline_taxonomy}(a) illustrates the end-to-end processing flow, starting from RF signal acquisition at ESC sensors, followed by preprocessing, feature extraction, and detection, ultimately leading to spectrum management decisions by the SAS. Complementarily, Fig.~\ref{fig:pipeline_taxonomy}(b) presents a taxonomy of detection approaches, highlighting the distinction between classical signal processing methods and modern machine-learning-based techniques. This combined view helps clarify how different algorithms fit within the overall ESC framework and emphasizes the evolution from model-driven to data-driven detection strategies.

\section{Classical Signal Processing Detection Methods}
\label{sec:Classical_Signal_Processing_Detection_Methods}

In early radar detection systems, researchers mainly relied on traditional signal processing techniques that analyze basic signal properties such as energy, structure, and repetition patterns. These approaches are generally simple to understand and implement, and they form the foundation of many modern detection systems. Each method, however, comes with its own strengths and limitations, especially when applied to challenging CBRS environments.

\noindent\textbf{Energy Detection:} Energy detection is the simplest approach, where the system measures the total energy of a received signal and compares it to a predefined threshold. If the energy exceeds this threshold, a radar signal is assumed to be present.
This method is attractive because it does not require any prior knowledge of the radar waveform, making it easy to deploy. However, it is highly sensitive to noise and interference. Signals from LTE or Wi-Fi can increase the measured energy and lead to incorrect detections. This limitation is commonly referred to as the \textit{SNR wall}~\cite{Shams2026}. As a result, energy detection
struggles to meet strict CBRS requirements, such as reliably detecting signals as weak as $-89$\,dBm/MHz with $99\%$ accuracy.

\noindent\textbf{Matched-Filter Detection:}
Matched filtering improves detection performance by comparing the received signal with a known radar waveform template. When the incoming signal closely matches the template, the system declares a detection. This method is highly effective and can detect very weak signals, even in the presence of interference, and has been shown to meet CBRS performance requirements~\cite{Caromi2018}.
However, its main limitation is the need for prior knowledge of the radar signal. If the waveform changes or is unknown, the detection performance can degrade significantly.

\noindent\textbf{Cyclostationary Feature Detection:}
Cyclostationary detection takes advantage of repeating patterns in radar signals, such as periodic pulses. By identifying these patterns, the method can distinguish radar signals from noise, which typically does not exhibit such structured behavior. This makes it more robust than energy detection, especially in noisy environments, and capable of detecting weaker signals~\cite{Shams2026}. Despite its advantages, this method is computationally intensive and requires more data and processing time, which can limit its use in real-time systems.

\noindent\textbf{Constant False Alarm Rate (CFAR):}
CFAR enhances detection reliability by dynamically adjusting the detection threshold
based on the surrounding noise level rather than using a fixed value. This allows the
system to maintain a consistent false alarm rate even when noise conditions change
over time~\cite{Shams2026,CFARPaper2026}. For example, in a noisy environment, the
threshold increases to reduce false alarms. While CFAR improves adaptability compared
to simple energy detection, its performance still depends on accurate estimation of
the noise level.

\noindent\textbf{Feature-Based Detection:} Feature-based methods focus on identifying specific characteristics of radar signals, such as pulse width, waveform shape, or frequency patterns like chirps. These methods can achieve high performance when the radar signal characteristics are well known. However, they are less flexible and may not generalize well to unknown or dynamically changing radar signals.

Classical detection methods are well established and relatively easy to implement, but each method involves trade-offs. Energy detection is simple but unreliable in noisy environments, matched filtering is highly accurate but depends on known signal patterns, cyclostationary methods are robust but computationally expensive, CFAR provides adaptability but relies on accurate noise estimation, and feature-based methods perform well only when signal characteristics are known. Overall, while these methods can work effectively in controlled conditions, they often face challenges in real-world CBRS environments with interference and varying signal conditions. This has motivated the development of modern machine learning approaches to achieve more robust and reliable detection performance. A detailed comparison of these methods in terms of complexity, robustness, and deployment suitability is provided in Table~\ref{tab:methods}.

\begin{table*}[!t]
  \centering
  \caption{Comparative Analysis of Radar Detection Methods in CBRS Sensing with Respect to Performance, Robustness, and Deployment Constraints.}
  \label{tab:methods}
  \renewcommand{\arraystretch}{1.2}
  \setlength{\tabcolsep}{3.5pt}
  \begin{tabular}{p{2.4cm}|p{1.3cm}|p{2.8cm}|p{3.0cm}|p{2.6cm}|p{2.2cm}}
    \hline
    \textbf{Method} & \textbf{Complexity} & \textbf{Knowledge / Data Requirement}
    & \textbf{Detection Performance} & \textbf{Robustness (Noise/Interference)} & \textbf{Latency Suitability} \\
    \hline
    
    Energy Detection 
    & Low 
    & None 
    & Moderate (80--90\%, degrades at low SNR) 
    & Poor (SNR wall, interference sensitive)~\cite{Shams2026} 
    & Excellent (real-time) \\
    
    \hline
    
    Matched Filter 
    & High 
    & Exact waveform knowledge 
    & Very High ($\approx 99\%$, optimal under match)~\cite{Caromi2018} 
    & Low (sensitive to mismatch/interference) 
    & Moderate (processing overhead) \\
    
    \hline
    
    Cyclostationary 
    & High 
    & Signal structure (PRF, periodicity) 
    & High (effective at low SNR) 
    & High (robust to noise uncertainty)~\cite{Shams2026} 
    & Low (computationally intensive) \\
    
    \hline
    
    CFAR (Adaptive ED) 
    & Medium 
    & Noise statistics estimation 
    & Adaptive (maintains $P_{fa}$ control) 
    & Moderate (depends on noise estimation)~\cite{CFARPaper2026} 
    & High (suitable for real-time) \\
    
    \hline
    
    Feature-Based ML 
    & Medium 
    & Handcrafted features + labeled data 
    & High (if features well-designed) 
    & Moderate (limited generalization) 
    & Moderate \\
    
    \hline
    
    CNN-based DL 
    & High 
    & Large labeled datasets (spectrogram/IQ) 
    & Very High ($>99\%$ at moderate/high SINR)~\cite{Lees2019,Caromi2021} 
    & High (handles interference and variability) 
    & Moderate (depends on hardware acceleration) \\
    
    \hline
    
    Lightweight ML/DL 
    & Low--Medium 
    & Reduced or transfer learning data 
    & Moderate to High (trade-off accuracy vs efficiency)~\cite{Khan2024} 
    & Moderate 
    & High (edge deployment feasible) \\
    
    \hline
    
    Hybrid (CFAR + DL) 
    & Medium--High 
    & Partial knowledge + labeled data 
    & Very High (near DL accuracy with lower cost) 
    & High (reduces false alarms + improves low SNR) 
    & High (optimized pipeline) \\
    
    \hline
  \end{tabular}
\end{table*}

\section{Machine Learning and Deep Learning Approaches}
\label{sec:Machine-Learning_and_Deep-Learning_Approaches}


In recent years, machine learning (ML) and deep learning (DL) have become popular tools for detecting radar signals in the CBRS band. Unlike traditional methods that rely on fixed rules, these approaches learn patterns directly from data, making them more flexible and better suited for complex real-world environments where signal conditions vary significantly.

\textbf{Supervised learning} is the most widely used approach in this domain. In this method, models are trained using labeled datasets where each example is identified as either ``radar present'' or ``no radar.'' Through this process, the model learns distinguishing patterns between the two classes. Earlier machine learning techniques relied on algorithms such as Support Vector Machines (SVMs)~\cite{8751641}, while more recent work has focused on deep learning models, particularly Convolutional Neural Networks (CNNs)~\cite{krizhevsky2012imagenet,he2016deep}. For instance, Lees \emph{et al.}~\cite{Lees2019} trained multiple models using approximately $14{,}000$ spectrogram images and demonstrated that deep learning significantly outperforms traditional methods. Their study also showed that even relatively small CNN architectures can provide a strong balance between detection accuracy and computational efficiency. Similarly, systems such as DeepRadar~\cite{DeepRadar} report detection accuracy close to $99\%$ under favorable signal conditions (e.g., $20$\,dB SIR), highlighting the ability of deep models to automatically extract important features such as pulse structures and frequency variations, even in the presence of interference~\cite{Caromi2021}. Despite these advantages, supervised learning methods require large labeled datasets, which are often expensive and time-consuming to obtain, and their performance may degrade when encountering signal patterns that differ significantly from the training data.

\textbf{Unsupervised learning and anomaly detection} provide an alternative approach when labeled data is limited or unavailable. Instead of learning from labeled examples, these methods model the characteristics of ``normal'' signals, such as LTE or $5$G transmissions, and identify deviations from this baseline as radar activity. Techniques such as clustering, autoencoders, and one-class SVMs are commonly used for this purpose. While these methods are promising, especially in dynamic environments, they are still under active research and have not yet been widely adopted in practical CBRS deployments.

To address data limitations, \textbf{transfer learning and data augmentation techniques} are increasingly being used. In transfer learning, a model is first trained on a large dataset, often synthetic, and then fine-tuned using a smaller set of real-world data~\cite{NISTDataset2020}. This approach reduces the need for extensive labeled datasets while still achieving good performance. Data augmentation further enhances model robustness by generating additional training samples through controlled modifications such as adding noise or varying signal parameters. These techniques help models generalize better across different operating conditions.

Another important direction is the development of \textbf{lightweight and edge-aware models} suitable for real-time deployment. In practical CBRS systems, detection must occur quickly and often on hardware with limited computational resources. To meet these constraints, researchers are designing efficient models that reduce memory and processing requirements. For example, YOLO-based architectures, originally developed for object detection in images, have been adapted for radar detection in spectrogram representations. The SenseORAN system~\cite{senseORAN} demonstrates this approach and achieves $100\%$ detection accuracy at SINR values greater than or equal to $12$\,dB while operating in real time. Additional techniques such as model compression and quantization are also used to reduce computational complexity without significantly affecting accuracy.

Overall, machine learning and deep learning approaches provide significant advantages over traditional detection techniques. They are capable of learning complex signal patterns directly from data, perform well in noisy and interference-heavy environments, and can achieve very high detection accuracy, often exceeding $99\%$. However, these methods also face challenges, including the need for large training datasets, higher computational requirements, and potential limitations in generalizing to unseen conditions. As a result, while ML and DL approaches are highly promising and often outperform classical methods in real-world scenarios~\cite{Lees2019,Khan2024}, ongoing research continues to focus on improving their efficiency, robustness, and adaptability for practical CBRS deployments.

\section{Hybrid Detection Approaches}
\label{sec:Hybrid_Detection_Approaches}

Recently, hybrid detection methods have emerged as a promising direction for radar detection in the CBRS band. These approaches combine the strengths of classical signal processing techniques with the adaptability of machine learning models to achieve both reliability and robustness in complex environments.

In a typical hybrid pipeline, classical methods such as energy detection or Constant False Alarm Rate (CFAR) are first used to identify candidate signal segments. These methods are computationally efficient and provide fast initial screening of the spectrum. The detected candidates are then passed to machine learning or deep learning models, such as Convolutional Neural Networks (CNNs), for refined classification. This two-stage approach reduces computational cost while improving detection accuracy, especially under low signal-to-noise ratio (SNR) conditions. One common hybrid design integrates CFAR with deep learning models. In this setup, CFAR dynamically adjusts detection thresholds to identify potential radar events, while a CNN processes corresponding spectrograms to confirm whether the detected signals are true radar emissions or false alarms caused by interference. This combination helps maintain a balance between sensitivity and false alarm control. Another variant is \textit{feature-assisted deep learning}, where domain-specific signal features (e.g., pulse width, pulse repetition frequency, or cyclostationary properties) are extracted using classical techniques and then provided as additional inputs to machine learning models. By incorporating expert knowledge into the learning process, these methods improve model interpretability and generalization, particularly when training data is limited.

Hybrid approaches offer several advantages. They leverage the theoretical reliability and low-latency operation of classical methods while benefiting from the pattern recognition capabilities of deep learning. As a result, they are well suited for real-world ESC deployments, where systems must operate under strict performance constraints, limited computational resources, and highly dynamic interference conditions.

Overall, hybrid detection represents a practical and increasingly important research direction in CBRS radar sensing. Future work is expected to further optimize the integration between model-driven and data-driven components, enabling more efficient, adaptive, and scalable detection systems.

\begin{tcolorbox}[insightbox]
Classical detection techniques, such as matched filtering, provide optimal performance when radar waveforms are known a priori; however, their effectiveness degrades significantly in heterogeneous CBRS environments characterized by diverse radar types and strong interference. In contrast, deep learning approaches offer greater robustness by learning complex signal patterns directly from data, enabling improved performance under variability and uncertainty. This advantage, however, comes at the cost of increased computational requirements and reliance on large, well-curated training datasets. Consequently, a fundamental trade-off arises between model-driven optimality and data-driven adaptability, representing a key challenge in the design of efficient and reliable ESC systems.
\end{tcolorbox}

\section{Datasets and Simulation/Emulation Platforms}
\label{sec:Datasets_and_Simulation}


To develop and evaluate radar detection systems, researchers need data that represents how radar and communication signals behave. This data is used both to train models (especially in machine learning) and to test how well detection methods perform. In general, data comes from two main sources: artificially generated (synthetic) data and real-world collected data.

One widely used resource is the NIST synthetic radar dataset~\cite{NISTDataset2020,Caromi2021}. Instead of recording real signals, this dataset is created using computer simulations that follow official testing guidelines defined by NTIA~\cite{NTIA2017}. It includes many variations of radar signals, such as different pulse widths, repetition rates, and noise conditions. Because it is generated programmatically, it can provide a large number of labeled examples (for instance, around $70{,}000$ samples), which is very useful for training machine learning models. In addition, NIST provides tools (such as MATLAB scripts) that allow researchers to generate signals and add controlled interference, making it easier to test algorithms under different scenarios. In contrast, real-world datasets are collected by measuring actual radar signals in operational environments. For example, NIST Technical Note TN~$1954$~\cite{Hale2017} includes recorded radar waveforms collected near coastal areas, where naval radars operate. These datasets often contain raw IQ data, which preserves detailed signal information such as amplitude and phase. Other measurement campaigns have recorded large numbers of radar pulses (e.g., more than $223{,}000$ pulses) in real environments~\cite{Caromi2019}. While these datasets are more realistic, they are harder to collect, may contain noise and interference, and are often limited in size or availability.

Another important category is over-the-air datasets, which are collected using devices such as Software Defined Radios (SDRs) in real wireless environments. These datasets include both radar signals and communication signals like $4$G or $5$G, making them useful for studying interference conditions. For example, projects like SenseORAN~\cite{senseORAN} and Spec-SCAN~\cite{10976013} provide labeled spectrogram data showing whether radar is present. These datasets better reflect real-world conditions but are still relatively limited compared to synthetic data.
\begin{figure}[!t]
    \centering
    \includegraphics[width=\columnwidth]{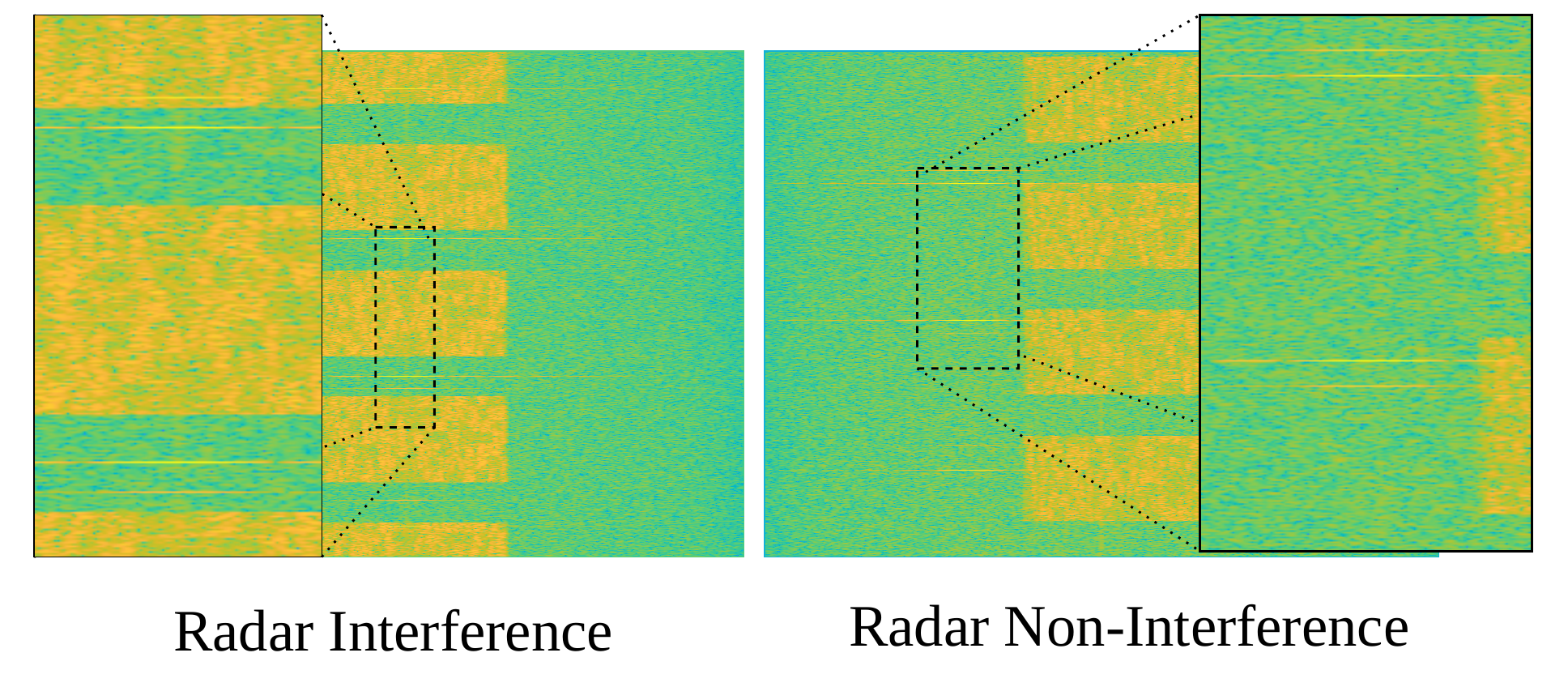}
    \caption{Time–frequency spectrograms of radar scenarios in the CBRS band. The left (interference) shows radar pulses overlapping with 4G/5G signals, creating mixed patterns that hinder detection. The right (non-interference) shows clean, periodic radar pulses without cellular overlap, highlighting the challenge of isolating radar signals in real-world deployments.}
    \label{fig:interference-non-interference}
\end{figure}
In addition to datasets, researchers use simulation and emulation platforms to test detection systems. Tools such as MATLAB and GNU Radio can simulate radar signals, wireless interference, and different channel conditions. NIST also provides simulation frameworks that model CBRS spectrum usage and radar activity~\cite{NISTDataset2020,Caromi2021}. These platforms allow controlled and repeatable experiments, which are useful for comparing different methods. However, simulations cannot fully capture real-world effects such as hardware imperfections or signal reflections, so they are often combined with real-world testing. An example of radar detection under interference and non-interference scenarios is illustrated in Fig.~\ref{fig:interference-non-interference}.

Overall, researchers typically use a combination of synthetic datasets (large and flexible), real measured data (accurate but limited), over-the-air recordings (realistic scenarios), and simulation tools (controlled experiments). Using all these sources together helps ensure that radar detection systems are reliable and effective in real-world CBRS deployments.

\section{Performance Metrics and Benchmarks}
\label{sec:Performance_Metrics_and_Benchmarks}

To evaluate the effectiveness of radar detection systems in the CBRS band, several performance metrics are used. The most fundamental metrics are the \textit{probability of detection} ($P_d$) and the \textit{probability of false alarm} ($P_{fa}$), which are formally defined under a binary hypothesis testing framework as:

\[
P_d = P(\text{Detection} \mid H_1), \qquad
P_{fa} = P(\text{Detection} \mid H_0)
\]

where $H_1$ represents the hypothesis that a radar signal is present, and $H_0$ represents the hypothesis that no radar signal is present. In this context, $P_d$ measures how often the system correctly detects radar signals, while $P_{fa}$ quantifies how often the system incorrectly declares radar presence in the absence of actual radar activity.

In CBRS systems, achieving a high detection probability is critical. The Federal Communications Commission (FCC) requires ESC systems to achieve $P_d \geq 0.99$ for very weak radar signals (as low as $-89$\,dBm/MHz) within a maximum detection time of $60$ seconds~\cite{FCC,FCC2015}. At the same time, maintaining a low false alarm rate is equally important, since excessive false alarms can unnecessarily disrupt commercial communications by triggering avoidable spectrum evacuation.

Another important interpretation of detection performance in modern systems, especially in machine learning-based approaches, is \textit{radar overlap recall}. This metric reflects the system’s ability to correctly identify time intervals where radar signals overlap with ongoing transmissions. In practice, radar overlap recall is closely related to $P_d$, but is evaluated over temporal segments or frames rather than individual decisions, making it particularly useful for real-time detection scenarios where signals are intermittent and time-varying. In addition to $P_d$ and $P_{fa}$, several complementary metrics are used to provide a more complete evaluation. \textit{Detection latency} measures the time required for the system to detect a radar signal after its appearance. Lower latency is desirable because it allows faster response and minimizes potential interference. In machine learning-based systems, standard classification metrics such as accuracy, precision, recall, and F1-score are also commonly reported to assess model performance. For example, SenseORAN~\cite{senseORAN} achieved $100\%$ detection accuracy at SINR values greater than or equal to $12$\,dB, demonstrating strong performance under favorable conditions.

Signal quality is another key factor influencing detection performance and is typically measured using the signal-to-interference-plus-noise ratio (SINR). FCC guidelines indicate that reliable detection ($P_d = 0.99$) should be achieved at SINR levels around $20$\,dB~\cite{FCC2015}. However, recent machine learning-based approaches have demonstrated the ability to operate effectively even at significantly lower SINR levels, in some cases as low as $-5$\,dB~\cite{Khan2024}, highlighting their robustness in challenging environments.

Finally, practical systems must also satisfy real-time operational constraints. Although the FCC allows up to $60$ seconds for detection, modern implementations aim for much faster response times. For instance, SenseORAN~\cite{senseORAN} achieves detection latency on the order of hundreds of milliseconds (approximately $866$ ms), enabling near real-time spectrum adaptation. Overall, these metrics collectively ensure that radar detection systems are not only accurate in identifying incumbent signals but also efficient, timely, and reliable under realistic CBRS operating conditions.

\section{Implementation and Deployment Challenges}
\label{sec:Implementation_and_Deployment_Challenges}

Building and using ESC systems in the real world is not just about detecting radar signals accurately; there are many practical challenges that must be handled carefully.

One major challenge is \textbf{false alarms}. A false alarm happens when the system thinks a radar signal is present when it is not. This can cause unnecessary shutdown of communication systems. Even though the FCC requires high detection probability, it does not strictly limit false alarm rates~\cite{FCC}. Because of this, system designers must carefully balance detecting real radar signals ($P_d$) while keeping false alarms ($P_{fa}$) low. In practice, systems often use conservative thresholds and require confirmation from multiple sensors to reduce mistakes. However, this can sometimes lead to missed detections if sensors are not well synchronized~\cite{Shams2026,CFARPaper2026}.

Another important issue is \textbf{detection latency}, which means how quickly the system can detect a radar signal. The FCC allows up to $60$ seconds for detection~\cite{FCC2015}, but faster detection is better in real systems. Some traditional methods need to observe signals over time before making a decision, while machine learning models may need to process chunks of data. Modern systems like SenseORAN process data in small time blocks (around $125$ milliseconds) to achieve very fast detection, sometimes in less than $1$ second~\cite{senseORAN}. This often requires powerful hardware such as FPGAs or specialized processors.

\textbf{Hardware and antenna design} is also a key challenge. ESC sensors must detect very weak signals (as low as $-89$\,dBm/MHz) while also handling strong nearby signals from LTE or $5$G networks~\cite{WInnForum1}. This requires high-quality radio components with wide bandwidth (around $150$ MHz) and strong filtering to remove unwanted signals~\cite{WInnForum2}. Sensors are often placed near coastlines and use directional antennas to focus on radar signals coming from the ocean. These systems must also be reliable and energy-efficient because they are often deployed in remote areas.

\textbf{Cooperative sensing} adds another layer of complexity. Instead of relying on a single sensor, multiple sensors work together to detect radar signals~\cite{WInnForum1}. This improves reliability because if one sensor misses a signal, another might detect it. However, combining data from multiple sensors requires careful coordination, time synchronization (often using GPS), and efficient communication between systems. It can also be difficult to determine whether multiple sensors are detecting the same radar event.

Another challenge is \textbf{interference affecting the sensors themselves}. ESC sensors must operate in environments where many wireless devices are active. Rules require that interference at the sensor should not exceed about $-109$\,dBm/MHz~\cite{WInnForum2}. To achieve this, special zones may be created around sensors where nearby devices reduce their transmission power. Managing this in a dynamic wireless network is difficult and requires coordination with the Spectrum Access System (SAS).

Finally, \textbf{cost and large-scale deployment} are major concerns. A complete ESC system may require hundreds of sensors along coastlines, each with installation, maintenance, and communication costs~\cite{WinForumSlides2024}. To reduce costs, some newer approaches try to reuse existing infrastructure, such as $5$G base stations acting as virtual sensors~\cite{senseORAN}. However, deploying and maintaining such a large network still requires significant investment and coordination.

In summary, real-world deployment of ESC systems involves balancing accuracy, speed, cost, and reliability. Even if a detection method works well in theory, it must also handle noise, interference, hardware limits, and large-scale coordination to be effective in practice.

\section{Spectrum Sharing Implications and Mitigation Strategies}
\label{sec:Spectrum_Sharing_Implications_and_Mitigation_Strategies}

In the CBRS system, radar detection is directly connected to how different users share the same spectrum. When a radar signal is detected, the Spectrum Access System (SAS) quickly adjusts or stops nearby commercial transmissions to protect the radar~\cite{FCC}. This means that accurate detection helps both sides: it protects important government systems while allowing commercial networks like LTE and $5$G to continue operating when it is safe. However, if the system makes mistakes, it can cause problems. For example, missing a radar signal can lead to interference with critical systems, while false alarms can unnecessarily shut down communication networks.

To reduce these problems, several mitigation strategies are used. One common approach is to quickly change how the spectrum is used when radar is detected. For example, if radar appears at a frequency like $3600$~MHz, nearby devices can switch to another available channel (such as a $20$~MHz block) or reduce their transmission power. In some cases, devices may temporarily stop transmitting until the radar signal disappears. More advanced systems use dynamic techniques, where networks can adapt in real time. For instance, newer architectures like O-RAN systems (e.g., SenseORAN~\cite{senseORAN}) can reconfigure network behavior in milliseconds, which is much faster than the traditional $60$-second requirement. This helps reduce interference more quickly and improves overall system efficiency.

Another important issue is interference from nearby frequency bands. The $3.45$--$3.55$~GHz band, which is close to the CBRS band, can create strong signals that interfere with ESC sensors~\cite{FCC2015}. To handle this, systems use filtering techniques and coordination rules to prevent signals from one band affecting another. Studies from WInnForum~\cite{WInnForum2} suggest that proper filtering and careful system design are necessary to ensure that ESC sensors can still detect radar signals accurately, even in crowded environments.

As more users and devices start using CBRS, especially indoor networks and private deployments, the level of interference in the spectrum increases. This makes detection more challenging and requires systems to be more robust. At the same time, improving radar detection has clear benefits. When detection is reliable, the system can allow more efficient use of the spectrum by reducing unnecessary protection areas. For example, if the system is confident in its detection, it can limit protection zones instead of blocking large regions, allowing more commercial users to operate.

Overall, effective spectrum sharing in CBRS depends on fast, accurate, and reliable radar detection. By using adaptive strategies such as dynamic frequency switching, power control, and real-time network reconfiguration, the system can balance protection of radar systems with efficient use of the spectrum for commercial applications.

\section{Open Research Problems and Future Directions}
\label{sec:Open_Research_Problems_and_Future_Directions}


Despite progress, many challenges and open problems remain in CBRS radar
detection.

\begin{itemize}[leftmargin=*]

\item \textbf{Low-SNR Detection.}
Achieving reliable detection at SINR well below 0~dB remains difficult.
Machine learning has pushed boundaries (e.g.,~detection at $-5$~dB
SINR~\cite{Khan2024}), but more work is needed to approach theoretical
limits. Techniques like data augmentation, synthetic data
generation~\cite{NISTDataset2020}, and joint waveform/channel estimation
could improve detection sensitivity.

\item \textbf{Adaptive and Transferable Models.}
Radar waveforms and environments change (different radars, seasonal
variations, new interference sources). Ensuring ML models generalize
across domains or can be quickly retrained (transfer learning, continual
learning) is crucial. Self-supervised or few-shot learning methods are
promising, building on advances for radar
classification~\cite{lees2019deep}.

\item \textbf{Real-Time Edge Implementation.}
Deploying ML detectors on resource-constrained sensors (or edge
processors) requires hardware-efficient models. Research on quantized
networks, neuromorphic hardware, or FPGA acceleration tailored to ESC is
needed.

\item \textbf{Cooperative/Networked Sensing.}
Optimal fusion of reports from multiple sensors remains an open problem.
Robust schemes that handle missing or delayed reports, or that
cross-verify radar detections, need exploration~\cite{WInnForum1}.
Graph-based ML techniques or distributed detection theory may play a role.

\item \textbf{Spectrum Mask and Privacy.}
ESC designs must comply with spectrum mask regulations~\cite{FCC} and
avoid any sensitive data collection. Research on anonymizing algorithms
that report only necessary metadata (time/freq) could ease deployment.

\item \textbf{Integration with 5G/6G Systems.}
Innovations like O-RAN, network slicing, and sensing-as-a-service open
new paradigms. For instance, embedding ESC functionality into 5G base
stations (as in SenseORAN~\cite{senseORAN}) can leverage massive MIMO and
network intelligence. How to standardize such architectures is an active
area.

\item \textbf{Adversarial Robustness and Security.}
As radar detection shifts toward deep learning, the vulnerability of these models to adversarial attacks becomes a concern. Malicious GAA users could theoretically transmit "pulse-like" interference designed to trigger a false positive in the ESC, effectively forcing a frequency handoff for competitors. Developing robust neural networks that can distinguish between intentional spoofing and actual naval radar is a critical security frontier.

\item \textbf{Multi-Band and International Harmonization.}
The 3.5~GHz band use is global, but rules vary~\cite{WInnForum2}.
Extending these methods to other shared bands (e.g.,~6~GHz Wi-Fi vs.\
incumbents, C-band satellite in 3.7~GHz) will require adapting detection
to different signals and regulatory regimes.

\end{itemize}

\section{Conclusions}
\label{sec:Conclusions}


Radar detection in the CBRS band is essential for allowing different users to safely share the same wireless spectrum. In this paper, we reviewed how the system works, including the rules set by regulators~\cite{FCC,FCC2015,WInnForum1,WInnForum2}, the types of radar signals that must be detected~\cite{Hale2017,Caromi2019}, and the main detection methods used today.

Traditional signal processing methods, such as matched filtering~\cite{Caromi2018}, work well when the radar signal is known and the environment is simple. However, in real-world conditions where signals are weak, noisy, or changing, these methods can struggle. On the other hand, modern machine learning approaches, especially deep learning models like CNNs~\cite{Lees2019,Caromi2021,DeepRadar}, have shown very strong performance, often achieving detection accuracy close to or above $99\%$. These models can learn complex patterns automatically, making them more effective in difficult environments. At the same time, deploying these systems in practice is not easy. ESC systems must detect very weak signals (as low as $-89$\,dBm/MHz) within a short time (up to $60$ seconds) while avoiding false alarms. They must also handle interference, hardware limitations, and coordination between multiple sensors. This makes real-world implementation more challenging than controlled experiments.

Overall, no single method is perfect. Simple methods are fast and easy to use but less accurate, while advanced learning-based methods are powerful but require more data and computation. Because of this, the most effective future systems will likely combine both approaches. For example, a system may first use simple signal processing to detect possible signals and then apply machine learning to confirm the result. In conclusion, CBRS radar detection is a rapidly evolving field. Future improvements should focus on making detection systems faster, more reliable, and easier to deploy in real-world environments. Achieving this balance will help protect important radar systems while allowing more efficient use of the shared spectrum for modern wireless communication.
\bibliographystyle{IEEEtran}
\bibliography{ref}

@misc{FCC,
  author       = {{Federal Communications Commission}},
  title        = {Citizens Broadband Radio Service},
  howpublished = {47~{CFR} Part~96, Subpart~G},
  note         = {\url{https://www.ecfr.gov/current/title-47/chapter-I/subchapter-D/part-96}}
}

@misc{FCC2015,
  author       = {{Federal Communications Commission}},
  title        = {Amendment of the {Commission's} Rules with Regard to
                  Commercial Operations in the 3550--3650~{MHz} Band},
  howpublished = {{GN} Docket No.~12-354, Report and Order},
  month        = apr,
  year         = {2015}
}

@techreport{WInnForum1,
  author      = {{Wireless Innovation Forum}},
  title       = {3.5~{GHz} Environmental Sensing Capability Detection
                 and Placement},
  institution = {Wireless Innovation Forum},
  number      = {WINNF-20-IN-0065},
  year        = {2020}
}

@techreport{WInnForum2,
  author      = {{Wireless Innovation Forum}},
  title       = {3.45~{GHz} {ESC} Coexistence},
  institution = {Wireless Innovation Forum},
  number      = {WINNF-RC-1016},
  year        = {2021}
}

@misc{WinForumSlides2024,
  author       = {{Wireless Innovation Forum}},
  title        = {Lower 3~{GHz} {ESC} Webinar},
  howpublished = {Presentation Slides},
  year         = {2024}
}

@techreport{Hale2017,
  author      = {Hale, Paul D. and Jargon, Jeffrey A. and Jeavons, Peter J.
                 and Souryal, Michael R. and Wunderlich, Adam J.
                 and Lofquist, Mark},
  title       = {3.5~{GHz} Radar Waveform Capture at Point {Loma,} {San Diego,
                 California}},
  institution = {National Institute of Standards and Technology ({NIST})},
  type        = {Technical Note},
  number      = {1954},
  year        = {2017},
  doi         = {10.6028/NIST.TN.1954}
}

@misc{NISTDataset2020,
  author       = {Caromi, Raied and Kallas, Kenneth and others},
  title        = {Simulated Radar Waveform and {RF} Dataset Generator
                  for 3.5~{GHz} {CBRS}},
  howpublished = {{NIST} National Research Data Repository},
  year         = {2020},
  doi          = {10.18434/mds2-2380}
}

@techreport{NTIA2017,
  author      = {Sanders, Frank H. and others},
  title       = {Procedures for Laboratory Testing of Environmental
                 Sensing Capability Sensor Devices},
  institution = {National Telecommunications and Information
                 Administration ({NTIA})},
  type        = {Technical Memorandum},
  number      = {TM-18-527},
  year        = {2017}
}

@inproceedings{Caromi2018,
  author    = {Caromi, Raied and Souryal, Michael R. and Yang, Wen-Bin},
  title     = {Detection of Incumbent Radar in the 3.5~{GHz} {CBRS} Band},
  booktitle = {Proc. {IEEE} Global Conference on Signal and Information
               Processing ({GlobalSIP})},
  address   = {Anaheim, CA, USA},
  month     = nov,
  year      = {2018},
  pages     = {241--245},
  doi       = {10.1109/GlobalSIP.2018.8646580}
}

@article{Caromi2019,
  author  = {Caromi, Raied and Souryal, Michael R. and Hall, Timothy A.},
  title   = {{RF} Dataset of Incumbent Radar Signals in the 3.5~{GHz}
             {CBRS} Band},
  journal = {Journal of Research of the National Institute of Standards
             and Technology},
  volume  = {124},
  pages   = {1--10},
  year    = {2019},
  doi     = {10.6028/jres.124.002}
}

@inproceedings{8751641,
  author    = {Caromi, Raied and Souryal, Michael R.},
  title     = {Detection of Incumbent Radar in the 3.5~{GHz} {CBRS}
               Band Using Support Vector Machines},
  booktitle = {Proc. Sensor Signal Processing for Defence ({SSPD})},
  address   = {Brighton, UK},
  year      = {2019},
  pages     = {1--5},
  doi       = {10.1109/SSPD.2019.8751641}
}

@inproceedings{Caromi2021,
  author    = {Caromi, Raied and Lackpour, Amir and Kallas, Kenneth
               and Nguyen, Thao and Souryal, Michael R.},
  title     = {Deep Learning for Radar Signal Detection in the 3.5~{GHz}
               {CBRS} Band},
  booktitle = {Proc. {IEEE} International Symposium on Dynamic Spectrum
               Access Networks ({DySPAN})},
  year      = {2021},
  pages     = {1--8},
  doi       = {10.1109/DySPAN53946.2021.9677280}
}

@article{Lees2019,
  author  = {Lees, William M. and Wunderlich, Adam and Jeavons, Peter J.
             and Hale, Paul D. and Souryal, Michael R.},
  title   = {Deep Learning Classification of 3.5-{GHz} Band Spectrograms
             With Applications to Spectrum Sensing},
  journal = {{IEEE} Transactions on Cognitive Communications and
             Networking},
  volume  = {5},
  number  = {2},
  pages   = {224--236},
  month   = jun,
  year    = {2019},
  doi     = {10.1109/TCCN.2019.2899871}
}

@article{lees2019deep,
  author  = {Lees, William M. and Wunderlich, Adam and Jeavons, Peter J.
             and Hale, Paul D. and Souryal, Michael R.},
  title   = {Deep Learning Classification of 3.5-{GHz} Band Spectrograms
             With Applications to Spectrum Sensing},
  journal = {{IEEE} Transactions on Cognitive Communications and
             Networking},
  volume  = {5},
  number  = {2},
  pages   = {224--236},
  month   = jun,
  year    = {2019},
  doi     = {10.1109/TCCN.2019.2899871}
}

@inproceedings{DeepRadar,
  author    = {Sarkar, Shamik and Buddhikot, Milind M. and Baset, Aitana
               and Kasera, Sneha Kumar},
  title     = {{DeepRadar}: A Deep-Learning-Based Environmental Sensing
               Capability Sensor Design for {CBRS}},
  booktitle = {Proc. {ACM} International Conference on Mobile Computing
               and Networking ({MobiCom})},
  year      = {2021},
  pages     = {56--68},
  doi       = {10.1145/3447993.3483259}
}

@inproceedings{soltani2022finding,
  author    = {Soltani, Nasim and Chaudhary, Vini and Roy, Debashri
               and Chowdhury, Kaushik R.},
  title     = {Finding {Waldo} in the {CBRS} Band: Signal Detection and
               Localization in the 3.5~{GHz} Spectrum},
  booktitle = {Proc. {IEEE} Global Communications Conference ({GLOBECOM})},
  year      = {2022},
  pages     = {4570--4575},
  doi       = {10.1109/GLOBECOM48099.2022.10000956}
}

@article{ReusMuns2023,
  author  = {Reus-Muns, Guillem and Upadhyaya, Pratheek S. and Demir, Ufuk
             and Stephenson, Nathan and Soltani, Nasim and Shah, Vijay K.
             and Chowdhury, Kaushik R.},
  title   = {{SenseORAN}: {O-RAN}-Based Radar Detection in the {CBRS} Band},
  journal = {{IEEE} Journal on Selected Areas in Communications},
  volume  = {42},
  number  = {2},
  pages   = {326--338},
  month   = feb,
  year    = {2024},
  doi     = {10.1109/JSAC.2023.3336162}
}

@article{senseORAN,
  author  = {Reus-Muns, Guillem and Upadhyaya, Pratheek S. and Demir, Ufuk
             and Stephenson, Nathan and Soltani, Nasim and Shah, Vijay K.
             and Chowdhury, Kaushik R.},
  title   = {{SenseORAN}: {O-RAN}-Based Radar Detection in the {CBRS} Band},
  journal = {{IEEE} Journal on Selected Areas in Communications},
  volume  = {42},
  number  = {2},
  pages   = {326--338},
  month   = feb,
  year    = {2024},
  doi     = {10.1109/JSAC.2023.3336162}
}

@inproceedings{10976013,
  author    = {Hazari, Raju and Singh, Gaurav and Renjith, Devika
               and Krishnan, Divya and {Pavanitha B} and Olufowobi, Habeeb
               and Roy, Debashri},
  title     = {{Spec-SCAN}: Spectrum Learning in Shared Channel Using
               Neural Networks},
  booktitle = {Proc. {IEEE} 22nd Consumer Communications \& Networking
               Conference ({CCNC})},
  month     = jan,
  year      = {2025},
  pages     = {1--6},
  doi       = {10.1109/CCNC56845.2025.10976013}
}

@misc{Khan2024,
  author        = {Khan, Shafi Ullah and Kulhandjian, Michel and Roy, Debashri},
  title         = {Pushing the Boundaries in {CBRS} Band: Robust Radar
                   Detection within High {5G} Interference},
  howpublished  = {arXiv preprint arXiv:2510.10040},
  year          = {2025},
  url           = {https://arxiv.org/abs/2510.10040}
}

@inproceedings{Shafi2025,
  author    = {Khan, Shafi Ullah and Kulhandjian, Michel and Roy, Debashri},
  title     = {In-Network Fusion for High Interference Signal Detection
               within {CBRS} Band},
  booktitle = {Proc. {IEEE} International Conference on Computer
               Communications ({INFOCOM})},
  month     = may,
  year      = {2025}
}

@misc{Shams2026,
  author       = {Shams, M. S. and Abouelfadl, A. A. and others},
  title        = {Adaptive Frequency-Domain {CFAR} for Robust Spectrum
                  Sensing},
  howpublished = {Research Square Preprint},
  year         = {2026}
}

@misc{CFARPaper2026,
  author       = {Shams, M. S. and others},
  title        = {Adaptive Frequency-Domain {CFAR}},
  howpublished = {Research Square Preprint},
  year         = {2026}
}

@inproceedings{krizhevsky2012imagenet,
  author    = {Krizhevsky, Alex and Sutskever, Ilya and Hinton, Geoffrey E.},
  title     = {{ImageNet} Classification with Deep Convolutional Neural
               Networks},
  booktitle = {Advances in Neural Information Processing Systems
               ({NeurIPS})},
  volume    = {25},
  pages     = {1097--1105},
  year      = {2012}
}

@inproceedings{he2016deep,
  author    = {He, Kaiming and Zhang, Xiangyu and Ren, Shaoqing and Sun, Jian},
  title     = {Deep Residual Learning for Image Recognition},
  booktitle = {Proc. {IEEE} Conference on Computer Vision and Pattern
               Recognition ({CVPR})},
  year      = {2016},
  pages     = {770--778},
  doi       = {10.1109/CVPR.2016.90}
}

\end{document}